\documentclass[twocolumn]{aastex6}
\usepackage{amssymb, amsmath,framed}

\usepackage{bm}
\expandafter\ifx\csname package@font\endcsname\relax\else
 \expandafter\expandafter
 \expandafter\usepackage
 \expandafter\expandafter
 \expandafter{\csname package@font\endcsname}
\fi
\def\bq{\begin{equation}}
\def\eq{\end{equation}}
\def\bqy{\begin{eqnarray}}
\def\eqy{\end{eqnarray}}

\def\ep{\epsilon}

\begin{document}

\title[Extended MHD turbulence]{Extended MHD turbulence and its applications to the solar wind}
\author{Hamdi M. Abdelhamid}
\email{hamdi@ppl.k.u-tokyo.ac.jp}
\affil{Graduate School of Frontier Sciences, University of Tokyo, Kashiwanoha, Kashiwa, Chiba 277-8561, Japan\\
Physics Department, Faculty of Science, Mansoura University, Mansoura 35516, Egypt}
\author{Manasvi Lingam}
\email{mlingam@princeton.edu}
\affil{Department of Astrophysical Sciences, Princeton University, Princeton, NJ 08544, USA}
\author{Swadesh M. Mahajan}
\affil{Department of Physics and Institute for Fusion Studies, The University of Texas at Austin, Austin, TX 78712, USA\\
Department of Physics, School of Natural Sciences, Shiv Nadar University, Uttar Pradesh 201314, India}

\begin{abstract}
Extended MHD is a one-fluid model that incorporates two-fluid effects such as electron inertia and the Hall drift. This model is used to construct fully \emph{nonlinear} Alfv\'enic wave solutions, and thereby derive the kinetic and magnetic spectra by resorting to a Kolmogorov-like hypothesis based on the constant cascading rates of the energy and generalized helicities of this model. The magnetic and kinetic spectra are derived in the ideal $\left(k < 1/\lambda_i\right)$, Hall $\left(1/\lambda_i < k < 1/\lambda_e \right)$, and electron inertia $\left(k > 1/\lambda_e\right)$ regimes; $k$ is the wavenumber and $\lambda_s = c/\omega_{p s}$ is the skin depth of species `$s$'. In the Hall regime, it is shown that the emergent results are fully consistent with previous numerical and analytical studies, especially in the context of the solar wind. The focus is primarily on the electron inertia regime, where magnetic energy spectra with power-law indexes of $-11/3$ and $-13/3$ are always recovered. The latter, in particular, is quite close to recent observational evidence from the solar wind with a potential slope of approximately $-4$ in this regime. It is thus plausible that these spectra may constitute a part of the (extended) inertial range, as opposed to the standard `dissipation' range paradigm.
\end{abstract}

\maketitle


\section{Introduction} \label{SecIntro}
In the realm of plasma astrophysics, a great deal of attention has been centred on ideal MHD, as it is the simplest of the plasma fluid models. Despite its considerable generality and elegance, ideal MHD is valid only in certain regimes, when a certain set of conditions are valid \citep{HW04}. In standard plasma physics texts, ideal MHD is often derived as a limiting case of the two-fluid model, the latter of which is obtained by taking moments of the Boltzmann equation \citep{KT73,HW04}. When two-fluid theory is expressed in terms of one-fluid variables (the centre-of-mass velocity and the current), along with some concomitant simplifications \citep{HW04}, the ensuing result is extended MHD.

Extended MHD is endowed with two chief two-fluid effects: (i) the Hall drift (as the electron and ion fluid velocities are different), and (ii) electron inertia (stemming from the finite, but small, mass of electrons). When both of these effects are neglected, ideal MHD is recovered in this limit, whilst Hall MHD is obtained when the electrons are assumed to be massless. Extended MHD has been known at least since the 1950s \citep{Spi56,Lust59}, although it has been extensively studied from a theoretical perspective only quite recently. Extended MHD effects typically become increasingly important as one approaches smaller length scales. In particular, the ideal MHD regime is characterized by $L > \lambda_i$, the Hall regime entails $\lambda_i > L > \lambda_e$, and extended MHD (with electron inertia) is valid for $L < \lambda_e$; note that $\lambda_s = c/\omega_{ps}$ is the skin depth of species `$s$' and $L$ is the scale length. It is important to recognize that extended MHD is a much more encompassing model than ideal MHD, but it does not capture any kinetic behaviour (such as Landau damping) or dissipative effects (for e.g., viscosity and resistivity).

Having said that, extended MHD has still proven to be highly useful in several contexts. Even Hall MHD, the simplest version of extended MHD, has been successfully employed in many contexts ranging from neutron stars \citep{CAZ04} and protoplanetary discs \citep{Ward07} to turbulence, outflows and dynamos \citep{MGM02,MGM03,MSMS05,LKF14,LM15,LB16,LinB16}. In the past two decades, Hall MHD has been applied to the solar wind with a fair degree of success, as evident from the (representative) studies undertaken by \citet{GSRG96,GG97,KM04,HFOZ05,GB07,ACVS07,ACVS08,SMC08,SS09,MG12,MA14,SP15}. The success of Hall MHD in space and laboratory plasmas also deserves to be mentioned \citep{Huba95,BMTG16}, especially in the realm of magnetic reconnection \citep{Bisk00}.

Current measurements and analysis of the solar wind spectrum in the regime $L \lesssim \lambda_e$ (sometimes interpreted as the `dissipation' range) appear to suggest that a power law behaviour, with a slope of approximately $-4$, is manifested \citep{SHVL06,SGRK09,Set11,GWPS15}; see also \citet{SGBCR10}. However, the measurements in the dissipation range are prone to instrumentation errors, as pointed out in \citet{Set13}, which has also led to other interpretations of the spectrum in its vicinity \citep{Alex09,ALM12}. Moreover, at these small scales, the magnetic fluctuations are not purely homogeneous and exhibit signs of intermittency \citep{PGDS12}. On account of all the complexities inherent in solar wind turbulence, gaining a thorough understanding of this phenomenon is, arguably, one of the current major goals \citep{Gold01,BC13}.

It has become increasingly common to model the solar wind spectra at scales smaller than the ion (or electron) skin depth by means of (gyro)kinetic simulations \citep{Het08,Het11,CB11,SBG12,THD13} or hybrid fluid-kinetic models \citep{CDQB11,VMMM12,SVCV12,Pet13,CCJTR16}, but computational and analytic studies of the solar wind by means of Hall MHD are also quite common \citep{SS09,MG12,MA14,SP15}. As we have pointed out earlier, it is incorrect to use Hall MHD to study the physics near the electron skin depth (which equals the electron gyroradius when the electron plasma beta is around unity). For this reason, there have been several studies centred around electron MHD, which can include the effects of electron inertia \citep{BSD96,BSZCD99,DDKD00,NBGG03,Gal08,MG10}. However, a chief limitation of electron MHD is the assumption of stationary ions. As a result, the model cannot be applied to systems where the mean velocity is significant. 

As our model (extended MHD) is endowed with a mean flow, electron inertia and the Hall drift, it gives rise to both electron and Hall MHD as limiting cases \citep{KLMWW14}. For this reason, we shall employ it as our basic physical model in determining the energy and helicities spectra. Our method relies upon the derivation of fully nonlinear Alfv\'en wave solutions for extended MHD, which are then used for computing the spectra. We demonstrate that our model reproduces many previous results, both experimental and theoretical; on the latter front, we show that it yields spectra that are distinct from those predicted by Hall MHD, and that are quite similar to the observational data from the solar wind \citep{SGRK09,BC13} and other collisionless plasmas \citep{Lea98}.

The outline of the paper is as follows. In Secs. \ref{XMHDPrelim} and \ref{SecNSXMHD}, we present the mathematical preliminaries and nonlinear wave solutions of extended MHD. In Secs. \ref{SecSpecDis} and \ref{SecESRes} we compute the spectra of the extended MHD invariants and describe the various limiting cases. We follow this up with a detailed discussion, analysis and comparison of our results in Sec. \ref{SecDiscAn}. We conclude by summarizing our results in Sec. \ref{SecConc}.

\section{Extended MHD: the mathematical preliminaries} \label{XMHDPrelim}
In this Section, we present a brief overview of extended MHD, and discuss some of its chief mathematical properties.

The equations of extended MHD comprise of the continuity equation
\begin{eqnarray}
\label{conteq}
\frac{\partial \rho}{\partial t}=-\nabla\cdot\left(\rho\textbf{V}\right),
\end{eqnarray}
the dynamical equation for the velocity,
\begin{eqnarray}
\label{momeq}
\frac{\partial \textbf{V}}{\partial t}=-\left(\nabla\times\textbf{V}\right)\times\textbf{V}+\rho^{-1} \left(\nabla\times\textbf{B}\right)\times\textbf{B}^{\ast}\nonumber \\ -\nabla\left(h+\frac{V^{2}}{2}+d^{2}_{e}\frac{\left(\nabla\times\textbf{B}\right)^{2}}{2\rho^{2}}\right),
\end{eqnarray}
and the extended MHD Ohm's law
\begin{eqnarray}
\label{OhmXMHD}
\frac{\partial \textbf{B}^{\ast}}{\partial t}=\nabla\times\left(\textbf{V}\times\textbf{B}^{\ast}\right)- \nabla\times\left(\rho^{-1} \left(\nabla\times\textbf{B}\right)\times\textbf{B}^{\ast}\right)\nonumber\\+d^{2}_{e} \nabla\times\left(\rho^{-1} \left(\nabla\times\textbf{B}\right)\times\left(\nabla\times\textbf{V}\right)\right),
\end{eqnarray}
where
\begin{eqnarray}
\label{extB}
\textbf{B}^{\ast}=\textbf{B}+d^{2}_{e}\nabla\times\rho^{-1}\left(\nabla\times\textbf{B}\right),
\end{eqnarray}
is the suitable dynamical variable (instead of the conventional magnetic field), and is widely used in electron MHD \citep{GKR94}. In the above expressions, note that $h$ is the total enthalpy, and $d_e = c/\left(\omega_{pe} L\right) \equiv \lambda_e/L$ is the normalized electron skin depth, where $\omega_{pe}$ and $L$ are the electron plasma frequency and scale length of the system respectively. In our model $\rho$, ${\bf V}$, and $\textbf{B}$ denote the mass density, centre-of-mass velocity and the magnetic field respectively. It is important to recognize that these equations have been normalized in Alfv\'enic units, as discussed in \cite{AKY15}. Moreover, the length scales are normalized in units of the ion skin depth $\lambda_i = c/\omega_{pi}$ for the sake of simplicity, i.e. it amounts to setting $L = \lambda_i$. The incompressible limit of extended MHD is easily obtained by setting $\rho = 1$ in the normalized units. 

It is well known that extended MHD \citep{KLMWW14,AKY15,LMM16} yields a conserved energy of the form
\begin{eqnarray}
\label{Energy}
E&=&\int_{\Omega}\left\{\rho\left(\frac{\left|\bm{V}\right|^{2}}{2}+d_e^2 \frac{\left|\bm{J}\right|^{2}}{2\rho^2} + U\left(\rho\right)\right)+\frac{\left|\bm{B}\right|^{2}}{2}
\right\} d^{3}x, \nonumber \\
&=& \int_{\Omega}\left\{\rho\left(\frac{\left|\bm{V}\right|^{2}}{2} + U\left(\rho\right)\right)+\frac{\bm{B}\cdot\bm{B}^\ast}{2}
\right\} d^{3}x,
\end{eqnarray}
where $U$ is the internal energy (per unit mass) of the system. Note that the second term on the RHS of the first line arises from the electron kinetic energy. In Hall and ideal MHD, which treat the electrons as inertialess, this term is not present. Moreover, extended MHD is endowed with \emph{two} helicities akin to the magnetic or fluid helicity, given by
\begin{eqnarray}
\label{XMHDHel}
C_{\pm}&=&\int_{\Omega}\mathbf{P}_{\pm}^{\ast}\cdot\left(\nabla\times \mathbf{P}_{\pm}^{\ast}\right)\, d^{3}x ,
\end{eqnarray}
where $\mathbf{P}^{\ast}_{\pm}=\mathbf{V}+\theta_{\pm}\mathbf{A}^{\ast}$ and $\theta_{\pm}={\left(-1\pm\sqrt{1+4d^{2}_{e}}\right)}/{\left(2d^{2}_{e}\right)}$ constitute the two constants \citep{AKY15,LMM15}. It is important to understand that extended MHD (and Hall MHD) is different from ideal MHD in this respect, since the former has two invariants of the form (\ref{XMHDHel}) whilst ideal MHD has only one. This has to do with the fact that Hall MHD (and, in a similar manner, extended MHD as well) is a singular perturbation of ideal MHD \citep{MY98,YM99}.

\section{Nonlinear wave solutions of extended MHD} \label{SecNSXMHD}
In this Section, we shall derive a certain class of nonlinear wave solutions for incompressible extended MHD, and then study the various limiting cases of this solution. This is done by adopting an approach akin to the one outlined in \citet{KM04,MK05,MM09}. However, before proceeding further, we point out that an alternative path can be adopted - one that was delineated in \citet{AY16}. In this paper, the (noncanonical) Hamiltonian formulation is used, along with a relaxation principle along the lines of \citet{SI97,MY98} to obtain the solutions in the wave frame. This is followed by a Galilean boost to recover the wave solutions in the lab frame. We do not reproduce the details here, but the reader may consult \citet{AY16} for further details. 

\subsection{The derivation of the nonlinear wave solutions}
The equations of incompressible extended MHD can be manipulated, and thereby cast into the following form:
\begin{eqnarray}
\label{Vort1}
\frac{\partial \textbf{B}^{\ast}}{\partial t}=\nabla\times\big[\left(\textbf{V}-\nabla\times\textbf{B}\right)\times\textbf{B}^{\ast}\big],
\end{eqnarray}
\begin{eqnarray}
\label{Vort2}
\frac{\partial \left(\textbf{B}^{\ast}+\nabla\times\textbf{V}\right)}{\partial t}=\nabla\times\big[\textbf{V}\times\left(\textbf{B}^{\ast}+\nabla\times\textbf{V}\right)\big],
\end{eqnarray}
along with the auxiliary condition 
\begin{eqnarray}
\label{Align3}
\nabla\times\left( \left(\nabla\times\textbf{B}\right)\times\left(\nabla\times\textbf{V}\right)\right)=0,
\end{eqnarray}
which will be commented on later. For now, it suffices to note that this term will obviously vanish when the magnetic and velocity fields are parallel (or anti-parallel) to one another. Furthermore, the condition (\ref{Align3}) eliminates the last term on the RHS of (\ref{OhmXMHD}), and thereby enables us to arrive at (\ref{Vort1}) and (\ref{Vort2}). The above equations must be supplemented with the incompressibility conditions
\begin{eqnarray}
\label{Inc1}
&& \nabla\cdot\textbf{V}=0, \\
\label{Inc2}
&& \nabla\cdot\textbf{B}^{\ast}=0,\quad \quad \nabla\cdot\textbf{B}=0.
\end{eqnarray}
We shall now describe a class of nonlinear waves that were first derived and investigated in \citet{KM04,MK05,MM09}; the electron inertia corrections that arise are explicitly displayed throughout.

Assuming that there is no ambient flow, we can split the velocity and magnetic field into the ambient and wave components, denoted by the subscript `$\circ$' and lowercase letters respectively, as follows
\begin{eqnarray}
\label{ambi1}
\textbf{B}=\widehat{\textbf{e}}_{B_{\circ}}+\textbf{b},\quad \quad \textbf{V}=\textbf{v},
\end{eqnarray}
where $\widehat{\textbf{e}}_{B_{\circ}}$ is the direction that the constant ambient field (in the normalized units) is oriented. Using the definition (\ref{extB}), we find that
\begin{eqnarray}
\label{ambi2}
\textbf{B}^{\ast}=\widehat{\textbf{e}}_{B_{\circ}}+\textbf{b}^{\ast},\quad \quad \textbf{b}^{\ast}=\textbf{b}+d^{2}_{e}\nabla\times\left(\nabla\times\textbf{b}\right).
\end{eqnarray}
Substituting  (\ref{ambi1}) and (\ref{ambi2}) into (\ref{Vort1}) and (\ref{Vort2}), the resultant equations are
\begin{eqnarray}
\label{VW1}
\frac{\partial \textbf{b}^{\ast}}{\partial t}&=&\nabla\times\big[\left(\textbf{v}-\nabla\times\textbf{b}\right)\times\textbf{b}^{\ast}\big] \nonumber \\
&& +\, \nabla\times\big[\left(\textbf{v}-\nabla\times\textbf{b}\right)\times\widehat{\textbf{e}}_{B_{\circ}}\big] ,
\end{eqnarray}
\begin{eqnarray}
\label{VW2}
\frac{\partial \left(\textbf{b}^{\ast}+\nabla\times\textbf{v}\right)}{\partial t}&=& \nabla\times\big[\textbf{v}\times\left(\textbf{b}^{\ast}+\nabla\times\textbf{v}\right)\big] \nonumber \\
&& +\, \nabla\times\big[\textbf{v}\times\widehat{\textbf{e}}_{B_{\circ}}\big].
\end{eqnarray}
Let us now suppose that the following (special) conditions were to be satisfied
\begin{eqnarray}
\label{Bel1}
\textbf{b}^{\ast}=\frac{1}{\mu_{1}}\left(\textbf{v}-\nabla\times\textbf{b}\right),
\end{eqnarray}
\begin{eqnarray}
\label{Bel2}
\textbf{b}^{\ast}+\nabla\times\textbf{v}=\frac{1}{\mu_{2}}\textbf{v}.
\end{eqnarray}
By substituting (\ref{Bel1}) and (\ref{Bel2}) into (\ref{VW1}) and (\ref{VW2}), the nonlinear terms are eliminated, leaving us with the following linear time-dependent equations:
\begin{eqnarray}
\label{VWL1}
\frac{\partial \textbf{b}^{\ast}}{\partial t}=\mu_{1} \nabla\times\big[\textbf{b}^{\ast}\times\widehat{\textbf{e}}_{B_{\circ}}\big] ,
\end{eqnarray}
\begin{eqnarray}
\label{VWL2}
\frac{\partial\textbf{v}}{\partial t}=\mu_{2}\nabla\times\big[\textbf{v}\times\widehat{\textbf{e}}_{B_{\circ}}\big],
\end{eqnarray}
which can be easily solved as they possess wave solutions of the form
\begin{eqnarray}
\label{VWLSol1}
\textbf{b}^{\ast}=\textbf{b}^{\ast}_{k}~\exp\big[i\textbf{k}\cdot\textbf{x}+i\mu_{1} \left(\widehat{\textbf{e}}_{B_{\circ}}\cdot\textbf{k}\right) t\big],
\end{eqnarray}
\begin{eqnarray}
\label{VWLSol2}
\textbf{v}=\textbf{v}_{k}~\exp\big[i\textbf{k}\cdot\textbf{x}+i\mu_{2}\left(\widehat{\textbf{e}}_{B_{\circ}}\cdot\textbf{k}\right) t\big].
\end{eqnarray}
But, in addition to satisfying (\ref{VWL1}) and (\ref{VWL2}), they must also meet the additional constraints imposed by (\ref{Bel1}) and (\ref{Bel2}). This necessitates $\mu_1 = \mu_2 = \mu$, and transforms (\ref{Bel1}) and (\ref{Bel2}) into
\begin{eqnarray}
\label{WSBel1}
\textbf{v}_{k}-\mu\textbf{b}^{\ast}_{k}=i \textbf{k}\times\textbf{b}_{k},
\end{eqnarray}
\begin{eqnarray}
\label{WSBel2}
\textbf{v}_{k}-\mu\textbf{b}^{\ast}_{k}=i \mu\textbf{k}\times\textbf{v}_{k}.
\end{eqnarray}
These two equations imply that
\begin{eqnarray}
\label{bvRel}
\textbf{b}_{k}= \mu\textbf{v}_{k},
\end{eqnarray}
which is a powerful relation between the fluctuating (wave) components of the flow and the magnetic field. This compact expression, along with (\ref{ambi1}), (\ref{VWLSol1}) and (\ref{VWLSol2}) will be shown to yield nonlinear Alfv\'en wave solutions of incompressible extended MHD. It has also been verified via back-substitution into the extended MHD equations.

Here, we wish to reiterate an important fact. It is the imposition of (\ref{Bel1}) and (\ref{Bel2}) that enables us to successfully handle the nonlinear terms inherent in (\ref{VW1}) and (\ref{VW2}). Thus, it appears as though the subsequent derivation, as exemplified by (\ref{VWL1}) and (\ref{VWL2}), is akin to a standard linear wave analysis. However, this is not merely a linear treatment, as the relations (\ref{Bel1}) and (\ref{Bel2}), which were essential in ``eliminating'' the nonlinearities, are analyzed and addressed in the discussion preceding (\ref{WSBel1}) and (\ref{WSBel2}), and in the equations themselves. 

Thus, our analysis does take into account all nonlinear terms, which are necessary in any study of turbulence as the latter involves scale-to-scale coupling. We observe that our use of the conditions (\ref{Bel1}) and (\ref{Bel2}) to eliminate the nonlinear contributions is a well-established approach \citep{KM04,MM09,MahLing15,AY16}. In fact, a similar result was also derived in \citet{Sch09} (see Footnote \#30), and the general methodology behind these approaches can be traced to the classic text of \citet{Whit74}.

A remarkable feature of (\ref{bvRel}) is that it satisfies the condition (\ref{Align3}), which was one of the conditions that we'd imposed at the beginning of our analysis. As stated earlier, we refer the reader to \citet{AY16} (see also \citealt{MM09}) for an alternative derivation, that does not rely upon this additional constraint for obtaining (\ref{bvRel}).

Let us now use (\ref{bvRel}) along with the expression $\textbf{b}^{\ast}_{k}=\textbf{b}_{k}-d^{2}_{e} \textbf{k}\times\left(\textbf{k}\times\textbf{b}_{k}\right)$, which follows from (\ref{ambi2}) and (\ref{VWLSol1}). We substitute these relations into either (\ref{WSBel1}) or (\ref{WSBel2}). This leads us to
\begin{eqnarray}
\label{DBVrel}
\textbf{k}\times\left(\textbf{k}\times\textbf{v}_{k}\right)+ \frac{i}{d^{2}_{e}\mu}\textbf{k}\times\textbf{v}_{k}=\frac{\left(\mu^{2}-1\right)}{d^{2}_{e}\mu^{2}}\textbf{v}_{k},
\end{eqnarray}
which can be further simplified upon using a suitable vector calculus identity, and $\textbf{k}\cdot\textbf{v}_k=0$. Thus, we end up with 
\begin{eqnarray}
\label{DBVRelF}
\textbf{k}\times\textbf{v}_{k} =-i\alpha\left(k\right)\textbf{v}_{k},
\end{eqnarray}
where $\alpha\left(k\right)$ is given by
\begin{eqnarray}
\label{alphadef}
\alpha\left(k\right) =\frac{\left(1-\mu^{2}\right)}{\mu}-\mu d^{2}_{e} k^{2}.
\end{eqnarray}
It is worth remarking that (\ref{DBVRelF}) is the Fourier transformed version of the Beltrami equation $\left(\nabla\times\textbf{v}=\alpha\textbf{v}\right)$ with a non-constant $\alpha$. We are now in a position to compute the final relation for $\alpha$ - this is done by taking the cross product of (\ref{DBVRelF}) with $\textbf{k}$, and then using (\ref{DBVRelF}) once again. We find that
\begin{eqnarray} \label{alphakrel}
\alpha\left(k\right) =\pm k,
\end{eqnarray}
which can then be combined with (\ref{alphadef}) to solve for $\mu$. The resulting equation is a quadratic, which leads to two solutions for $\mu$ (denoted by $\mu_\pm$) that are given by
\begin{eqnarray}
\label{mu}
\mu_{\pm}\left(k\right) =\frac{1}{\left(1+d^{2}_{e}k^{2}\right)}\left[-\frac{k}{2}\pm\sqrt{\frac{k^{2}}{4}+\left(1+d^{2}_{e}k^{2}\right)}\right],
\end{eqnarray}
and let us focus on the simple case wherein $\textbf{k}=k~\widehat{\textbf{e}}_{z}$, following the path prescribed in \citet{KM04,MK05}. The frequency  $\omega=-\mu\left(\widehat{\textbf{e}}_{B_{\circ}}\cdot \textbf{k}\right)$ reads as 
\begin{eqnarray}
\label{disprel}
\omega_{\pm} =\frac{-k}{\left(1+d^{2}_{e}k^{2}\right)}\left[-\frac{k}{2}\pm\sqrt{\frac{k^{2}}{4}+\left(1+d^{2}_{e}k^{2}\right)}\right]\left(\widehat{\textbf{e}}_{B_{\circ}}\cdot\widehat{\textbf{e}}_{z}\right).
\end{eqnarray}
The final expressions for (\ref{mu}) and (\ref{disprel}) are quite akin to the analogous results obtained in \citet{MM09}, except for the fact that \citet{MM09} relied upon the usage of two-fluid theory (and the corresponding variables).

Before concluding this part, we point out that each value of $\mu$ give rise to two distinct fully nonlinear wave solutions that resemble the famous ABC equilibria. We refer the reader to \citet{AY16} for an in-depth discussion of this issue.

\subsection{On the limiting cases of the nonlinear waves}\label{SSecLimCase}
We shall now investigate the various regimes of interest, and list the expressions for (\ref{mu}) and (\ref{disprel}) accordingly.

\begin{enumerate}
\item First, consider the limit $k \ll 1$, which in \emph{dimensional} units is tantamount to stating that $k d_i \ll 1$. This indicates that we are operating in the ideal MHD domain, and we arrive at
  \begin{equation} \label{MHDReg}
\mu_{\pm}\rightarrow \pm1,~~~~~~~\omega_{\pm}\rightarrow \mp k \left(\widehat{\textbf{e}}_{B_{\circ}}\cdot\widehat{\textbf{e}}_{z}\right),
\end{equation}
which corresponds to the shear Alfv\'en waves of ideal MHD (that are co- and counter-propagating).

\item Next, consider the case where Hall effects are important, but electron inertia can still be neglected, i.e. the Hall regime. In this instance, the conditions (in the dimensionless units) are given by $k>1$ and $d^{2}_{e} k^{2}\ll1$. The dispersion relations reduce to
\begin{eqnarray}
&& \mu_{+}\rightarrow 1/k,~~~~~~~~~\omega_{+}\rightarrow -1\left(\widehat{\textbf{e}}_{B_{\circ}}\cdot\widehat{\textbf{e}}_{z}\right),\nonumber \\
&& \mu_{-}\rightarrow -k,~~~~~~~~~\omega_{-}\rightarrow k^{2}\left(\widehat{\textbf{e}}_{B_{\circ}}\cdot\widehat{\textbf{e}}_{z}\right),
\end{eqnarray}
implying that $\omega_{+}$ is the magnetosonic-cyclotron branch since $\omega_{+}$ is the ion gyrofrequency. On the other hand, $\omega_{-}$ is the shear-whistler mode, as seen from the dispersion relation \citep{MK05,MM09,AY16}.

We record important features of the Hall regime before moving on to the next case. As opposed to the ideal MHD regime, or the electron inertia one (discussed below), the Hall regime is bounded strictly from below \emph{and} above. As a consequence, the range is somewhat `narrow' and care must be taken when investigating it in greater detail. Secondly, it may appear as though the whistler mode $\omega_{-}$ is unbounded as it is proportional to $k^2$. However, this is incorrect since we have implicitly assumed that the inequality $d_e^2 k^2 \ll 1$ is applicable. In turn, this suggests that the whistler mode is rendered invalid when considering frequencies higher than the electron gyrofrequency. 

\item The third regime of interest is when electron inertia effects become important, even dominant. This regime requires that the conditions $k \gg 1$ and $d^{2}_{e} k^{2}\gg 1$ be met. In this instance, we find that
\begin{equation}
\mu_{\pm}\rightarrow \theta_{\pm}/k,~~~~~~~\omega_{\pm}\rightarrow -\theta_{\pm}\left(\widehat{\textbf{e}}_{B_{\circ}}\cdot\widehat{\textbf{e}}_{z}\right),
\end{equation}
where $\theta_{\pm}= \left(-1\pm\sqrt{1+4d^{2}_{e}}\right)/2d^{2}_{e}$. By substituting the relation $d_e^2 \ll 1$, in terms of the normalized variables, into the expression for $\theta_\pm$, we note that $\theta_{-}$ approximates the normalized electron  gyrofrequency, whilst $\theta_{+}$ approximates the normalized ion gyrofrequency. It is important to recognize that $\theta_\pm$ depends on the dimensionless electron skin depth, and thereby gives rise to a direct relationship between the electron skin depth and $\omega_\pm$, i.e. the ion or electron gyrofrequency (for $\theta_{+}$ and $\theta_{-}$ respectively). \\
Thus, the magnetosonic-cyclotron branch in the electron inertia regime approaches the same limit as its Hall counterpart; this is seen by comparing the expressions for $\omega_{+}$ in both cases. However, the dispersion relation for the whistler-mode branch in Hall MHD (given by $\omega_{-}$), is such that it would diverge for $k \rightarrow \infty$. The role of electron inertia in this case is to impose a strict upper bound (viz. the electron gyrofrequency) on the frequency attainable by this whistler-mode.
\end{enumerate}

\section{The spectral energy distributions for extended MHD}\label{SecSpecDis}
In this Section, we shall focus on deriving the energy spectra for extended MHD by invoking a Kolmogorov-like hypothesis regarding the energy and generalized helicity cascades. We follow this up with pictorial representations of the various spectra.

\subsection{Extended MHD invariants in Fourier space}
We shall list the primary invariants of extended MHD, and give their Fourier representations. The total energy in extended MHD is given by
\begin{eqnarray}
\label{E1}
E&=&\frac{1}{2}\int_{\Omega}\left\{\left|\bm{V}\right|^{2}+\bm{B}\cdot\bm{B}^{\ast}
\right\} d^{3}x\nonumber\\
&=&\frac{1}{2}\sum_{k}\left\{\left|\bm{v}_{k}\right|^{2}+\left(1+d^{2}_{e}k^{2}\right)\left|\bm{b}_{k}\right|^{2}\right\},
\end{eqnarray}
and the relation $\textbf{b}^{\ast}_{k}=\textbf{b}_{k}-d^{2}_{e} \textbf{k}\times\left(\textbf{k}\times\textbf{b}_{k}\right)=\left(1+d^{2}_{e}k^{2}\right)\bm{b}_{k}$ was used to simplify, and eventually obtain, the above expression. 

Next, we are free to consider the helicities of extended MHD, which are given by (\ref{XMHDHel}). However, before we do so, we shall present an alternative set of invariants, which are fully equivalent to (\ref{XMHDHel}) instead. The motivation behind this stems from the following considerations. It is well known that the cross helicity $\int_\Omega \bm{V} \cdot \bm{B}\,d^3x$ and magnetic helicity $\int_\Omega \bm{A} \cdot \bm{B}\,d^3x$ are invariants of ideal MHD \citep{Wol58a,Wol58b}. In extended MHD, we have seen that $\bm{B}^{\ast}$ serves as the dynamical variable in the place of $\bm{B}$. Thus, it is natural to seek the extended MHD invariants, which involve $\bm{B}^{\ast}$ in the place of $\bm{B}$, that resemble the cross and magnetic helicities of ideal MHD.

We introduce an invariant $H$ which can be constructed from (\ref{XMHDHel}) which can be viewed as somewhat analogous to the cross helicity of ideal MHD, albeit with an additional helicity contribution. It is given by
\begin{eqnarray}
\label{H}
H&=&\int_{\Omega}\left(\bm{V}-\frac{1}{2d^{2}_{e}}\bm{A}^{\ast}\right)\cdot \bm{B}^{\ast}\,d^{3}x,\nonumber\\
&=&\sum_{k}\left\{ \bm{v}_{k}-\frac{i \left(1+d^{2}_{e}k^{2}\right)}{2d^{2}_{e}k^{2}}\left(\bm{k}\times\bm{b}_{k}\right)\right\} \nonumber \\
&& \quad \quad \cdot \quad \bigg\{\left(1+d^{2}_{e}k^{2}\right) \bm{b}_{-k}\bigg\},
\end{eqnarray}
where we have used $\textbf{A}^{\ast}_{k}=\left(1+d^{2}_{e}k^{2}\right)\bm{A}_{k}$, along with the relations
\begin{eqnarray*}
&&\textbf{B}^{\ast} = \nabla\times\textbf{A}^{\ast},\\
&&\textbf{B} = \nabla\times\textbf{A}, \\
&&\bm{A}_{k} = \frac{i}{k^{2}} \bm{k}\times\bm{b}_{k},
\end{eqnarray*}
the last of which follows from (\ref{bvRel}), (\ref{DBVRelF}) and (\ref{alphakrel}) along with the use of suitable vector calculus identities. It is more accurate to envision (\ref{H}) as a combination of cross and magnetic helicities, but they involve $\bm{B}^{\ast}$ (instead of $\bm{B}$) owing to electron inertia effects.

The second invariant $G$ introduced below can be viewed as the generalization of the magnetic helicity of ideal MHD. This invariant consists of two terms, the first of which is the equivalent of the magnetic helicity (with $\bm{B}^{\ast}$ in place of $\bm{B}$) and the second can be viewed as a fluid helicity correction. It is defined as follows:
\begin{eqnarray}
\label{G}
G&=&\frac{1}{2}\int_{\Omega}\big[\bm{B}^{\ast}\cdot\bm{A}^{\ast}+d^{2}_{e}\bm{V}\cdot\left(\nabla\times\bm{V}\right)\big] d^{3}x,\nonumber\\
&=&\frac{1}{2}\sum_{k}\Bigg\{ \frac{i \left(1+d^{2}_{e}k^{2}\right)^{2}}{k^{2}}\left(\bm{k}\times\bm{b}_{k}\right)\cdot\bm{b}_{-k} \nonumber \\
&& \hspace{0.5 in} -i d^{2}_{e}\bm{v}_{k}\cdot\bm{k}\times\bm{v}_{-k}\Bigg\}.
\end{eqnarray}

Next, let us consider the two generalized helicities given by (\ref{XMHDHel}). They can be expressed as follows: 
\begin{eqnarray}
\label{C}
C_\pm&=& \frac{1}{2}\int_{\Omega}\left(\theta_\pm\textbf{A}^{\ast}+\textbf{V}\right)\cdot\left(\theta_\pm\textbf{B}^{\ast}+\nabla\times\textbf{V}\right)d^{3}x,
\nonumber\\
&=&\frac{1}{2}\sum_{k}\left\{ \frac{i\theta_{\pm} \left(1+d^{2}_{e}k^{2}\right)}{k^{2}}\left(\bm{k}\times\bm{b}_{k}\right)+\bm{v}_{k}\right\}\nonumber\\ &&\cdot~~~\bigg\{ \theta_{\pm}\left(1+d^{2}_{e}k^{2}\right)\bm{b}_{-k}-i\bm{k}\times\bm{v}_{-k}\bigg\}.
\end{eqnarray}
where 
\begin{equation}
\left|\theta_{\pm}\right|=\left|\frac{-1\pm\sqrt{1+4d^{2}_{e}}}{2d^{2}_{e}}\right|\approx 
\begin{cases}
    1,& \text{for}\,\, \theta_{+}\\
    \frac{1}{d^{2}_{e}},& \text{for}\,\, \theta_{-}       
\end{cases}
\end{equation} 
as noted in Sec. \ref{SSecLimCase}. Here, we wish to reiterate that (\ref{H}), (\ref{G}) and (\ref{C}) are not independent; the former two can be constructed from the latter and vice-versa. Indeed, it is more `natural' to regard (\ref{C}) as the invariant helicities of extended MHD, on account of their similarity with magnetic helicity in terms of their structure.

We have also verified that any linear combination of the Casimir invariants generates the same results presented in Sec. \ref{SecNSXMHD}; most notably, we find that the same dispersion relation (\ref{disprel}) is recovered. This confirms that we are free to use either (\ref{H}) and (\ref{G}) or (\ref{C}) without loss of generality.

\subsection{The derivation of the spectra for extended MHD}\label{SSecSDer}
We are now in a position to derive the various spectra of interest in the different regimes. To do so, we shall rely upon an approach based on the classical arguments presented by \citet{Kol41}. We shall adopt the notation and methodology outlined in \citet{KM04} henceforth. In quantitative terms, we assume that the (total) energy cascade rate is the product of the energy (\ref{E1}) and the inverse of the eddy turnover time $\tau$, the latter of which is given by $\tau = \left(k |v_k|\right)^{-1}$. Thus, the energy cascading rate, denoted by $\ep_{E}$, is evaluated to be
\begin{eqnarray}
\label{epsilonEx}
\ep_{E}&=&k |v_k| \Big(1+\mu^{2}\left(1+d^{2}_{e}k^{2}\right)\Big)\frac{|v^{2}_{k}|}{2},
\end{eqnarray}
and we had invoked (\ref{bvRel}) to express (\ref{E1}) purely in terms of $v_k$. We introduce the omnidirectional spectral function $W_{E}\left(k\right)$ that corresponds to the kinetic energy per unit mass per unit wave vector $k$, and is thus found to equal
\begin{eqnarray}
\label{WEx}
W_{E}\left(k\right)&=&\frac{|v^{2}_{k}|}{k}\\
&=& \left(2\ep_{E}\right)^{2/3} k^{-5/3} \Big[1+\mu^{2}\left(1+d^{2}_{e}k^{2}\right)\Big]^{-2/3}. \nonumber
\end{eqnarray}
Similarly, it is possible to define the omnidirectional spectral distribution function for the magnetic energy density $M_{E}\left(k\right)$. This can be related to $W_{E}\left(k\right)$ by means of (\ref{bvRel}), thereby yielding
\begin{eqnarray}
\label{MEx}
M_{E}\left(k\right)&=& \mu^{2}W_{E}\left(k\right)
\end{eqnarray}
A similar set of arguments, and scaling relations can be thus devised for the helicities. As pointed our earlier, only two of (\ref{H}), (\ref{G}) and (\ref{C}) are truly independent. For the sake of completeness, however, we shall present the scaling relations for all of these helicities.  

Let us start with \eqref{H} first. Assuming that the eddy turnover time is $\tau$ as before, we find that its cascading rate $\left(\ep_{H}\right)$ is
\begin{eqnarray}
\label{epsilonHx}
\ep_{H} = k|v_{k}|\left[\mu\left(1+d^{2}_{e}k^{2}\right)\Big(1-\mu\frac{\left(1+d^{2}_{e}k^{2}\right)}{2d^{2}_{e} k}\Big)\right]|v^{2}_{k}|.
\end{eqnarray}
The associated kinetic and magnetic spectral energy can be computed in a similar manner, and they are given by
\begin{eqnarray}
\label{WHx}
W_{H}\left(k\right)&=&  \left[\mu\left(1+d^{2}_{e}k^{2}\right)\Big(1-\mu\frac{\left(1+d^{2}_{e}k^{2}\right)}{2d^{2}_{e} k}\Big)\right]^{-2/3} \nonumber \\
&& \quad \times\,\left(\ep_{H}\right)^{2/3} k^{-5/3},
\\
\label{MHx}
M_{H}\left(k\right)&=& \mu^{2}W_{H}\left(k\right).
\end{eqnarray}

Next, we consider the helicity \eqref{G} and repeat the above set of arguments and algebra. The cascading rate $\ep_G$ takes on the form
\begin{eqnarray}
\label{epsilonGx}
\ep_{G}&=& k|v_{k}|\Big[d^{2}_{e}k^{2}+\mu^{2}\left(1+d^{2}_{e}k^{2}\right)^{2}\Big] \frac{|v^{2}_{k}|}{2k},
\end{eqnarray}
and the spectral energies are
\begin{eqnarray}
\label{WGx}
W_{G}\left(k\right)&=& \frac{\left(2\ep_{G}\right)^{2/3}}{k} \Big[d^{2}_{e}k^{2}+\mu^{2}\left(1+d^{2}_{e}k^{2}\right)^{2}\Big]^{-2/3},
\\
\label{MGx}
M_{G}\left(k\right)&=& \mu^{2}W_{G}\left(k\right).
\end{eqnarray}

Moving on the generalized helicities (which are arguably the true analogs of the magnetic or fluid helicity), we find that the cascading rate $\ep_{C_\pm}$ is
\begin{eqnarray}
\label{epsilonCx}
\ep_{C_\pm}&=& k|v_{k}| \Big(k+\theta_{\pm}\mu\left(1+d^{2}_{e}k^{2}\right)\Big)^{2} \frac{|v^{2}_{k}|}{2k},
\end{eqnarray}
leading us to the spectral energies
\begin{eqnarray}
\label{WCx}
W_{C_\pm}\left(k\right)&=& \frac{\left(2\ep_{C_\pm}\right)^{2/3}}{k} \Big[k+ \theta_{\pm}\mu\left(1+d^{2}_{e}k^{2}\right)\Big]^{-4/3}, \\
\label{MCx}
M_{C_\pm}\left(k\right)&=& \mu^{2}W_{C_\pm}\left(k\right).
\end{eqnarray}
We round off this section by pointing out the fact that there are \emph{two} different values of $\mu$ that are given by (\ref{mu}). Hence, for each of the spectral energies, the two cases must be considered separately.

\subsection{The kinetic and magnetic spectral plots} \label{SSecSpecPlot}
\begin{figure*}
$$
\begin{array}{ccc}
 \includegraphics[width=16.0cm]{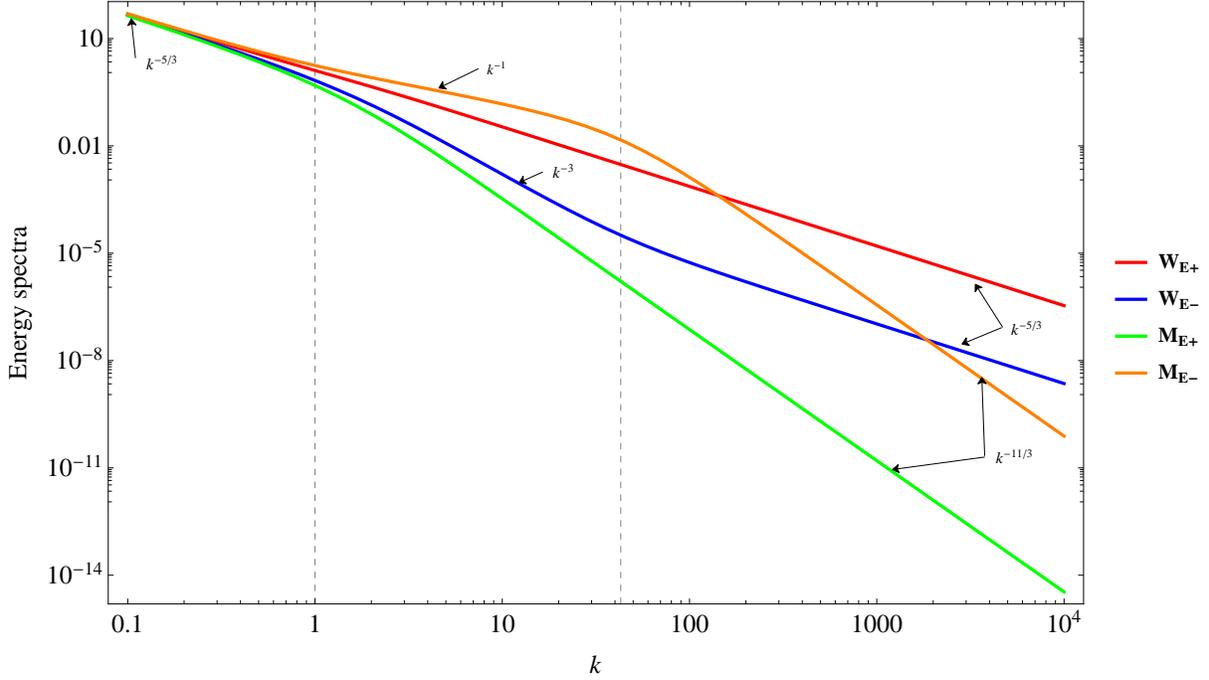}\\
\end{array}
$$
\caption{(colour figures online) Here, $W_{E_+}$ and $W_{E_-}$ are the two values of (\ref{WEx}) corresponding to $\mu_{+}$ and $\mu_{-}$ respectively; the latter duo are given by (\ref{mu}). Recall that $k$ has been normalized in units of $1/d_i$. The values of $M_{E_+}$ and $M_{E_-}$ are computed by means of (\ref{MEx}). The two vertical dotted lines separate the ideal, Hall and electron inertia regimes respectively (when viewed from left to right).}
\label{fig1}
\end{figure*}

\begin{figure*}
$$
\begin{array}{ccc}
 \includegraphics[width=16.0cm]{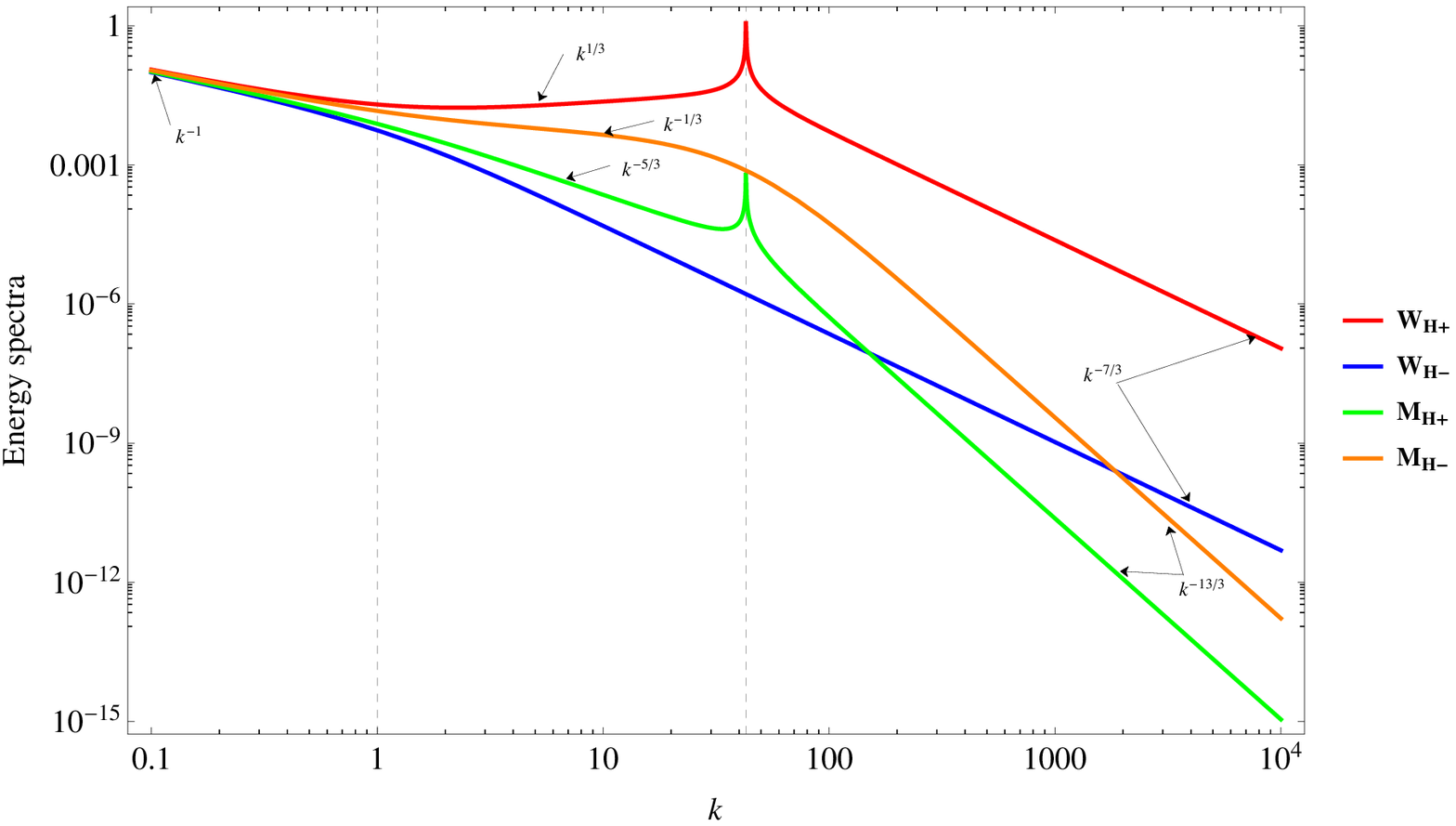}\\
\end{array}
$$
\caption{(colour figures online) Here, $W_{H_+}$ and $W_{H_-}$ are the two values of (\ref{WHx}) corresponding to $\mu_{+}$ and $\mu_{-}$ respectively; the latter duo are given by (\ref{mu}). Recall that $k$ has been normalized in units of $1/d_i$. The values of $M_{H_+}$ and $M_{H_-}$ are computed by means of (\ref{MHx}). The two vertical dotted lines separate the ideal, Hall and electron inertia regimes respectively (when viewed from left to right).}
\label{fig2}
\end{figure*}

\begin{figure*}
$$
\begin{array}{ccc}
 \includegraphics[width=16.0cm]{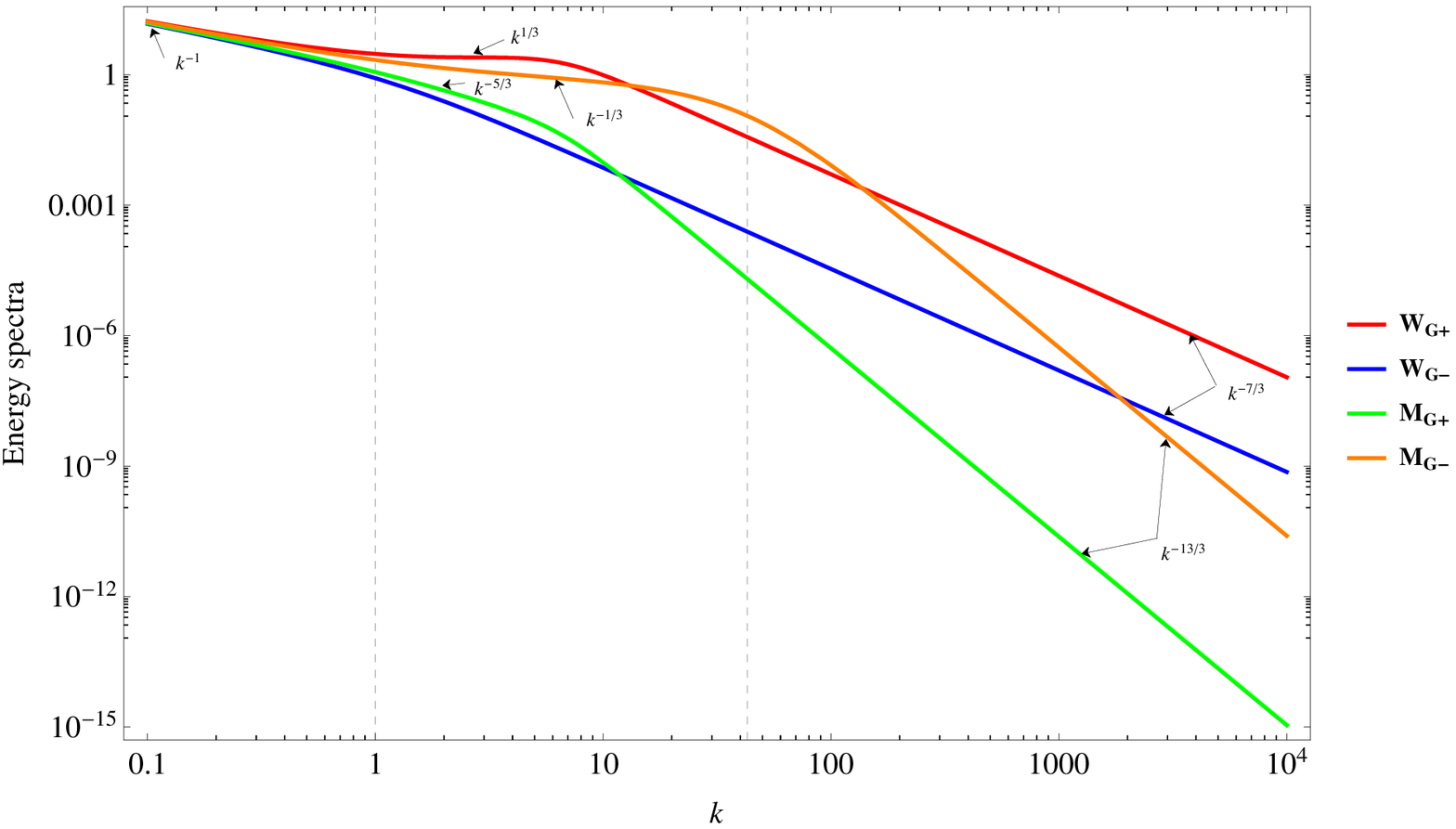}\\
\end{array}
$$
\caption{(colour figures online) Here, $W_{G_+}$ and $W_{G_-}$ are the two values of (\ref{WGx}) corresponding to $\mu_{+}$ and $\mu_{-}$ respectively; the latter duo are given by (\ref{mu}). Recall that $k$ has been normalized in units of $1/d_i$. The values of $M_{G_+}$ and $M_{G_-}$ are computed by means of (\ref{MGx}). The two vertical dotted lines separate the ideal, Hall and electron inertia regimes respectively (when viewed from left to right).}
\label{fig3}
\end{figure*}

\begin{figure*}
$$
\begin{array}{ccc}
 \includegraphics[width=16.0cm]{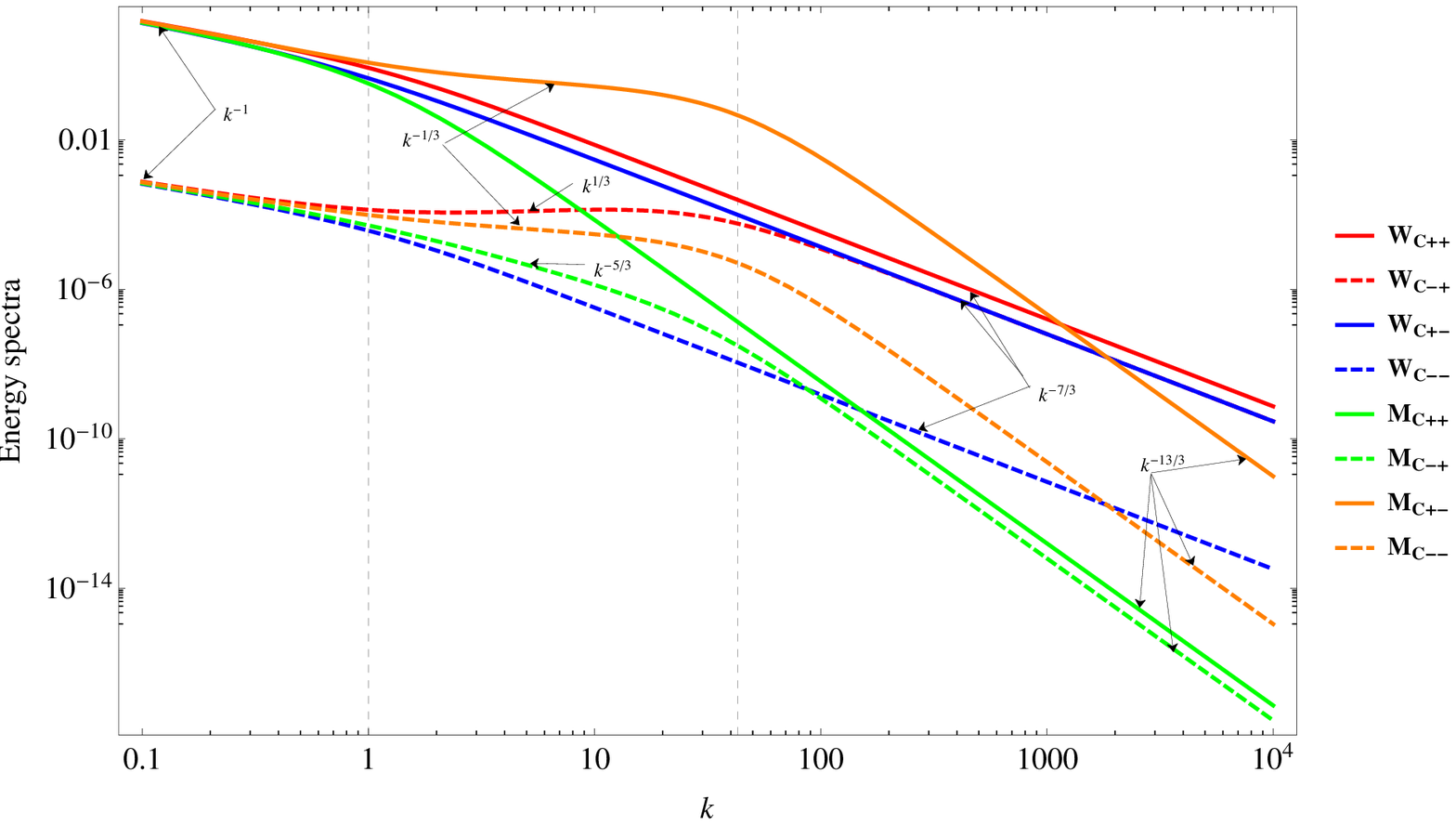}\\
\end{array}
$$
\caption{(colour figures online) Here, the $W_C$'s are the \emph{four} values of (\ref{WCx}) corresponding to $\mu_{+}$ and $\mu_{-}$ respectively. The first sign denotes the choice of $C$ (either $C_+$ and $C_-$) and the second denotes the choice of $\mu$, whose expressions are given by (\ref{mu}). Recall that $k$ has been normalized in units of $1/d_i$. The values of the $M_C$'s are found by using (\ref{MCx}), and they are also four in number. The two vertical dotted lines separate the ideal, Hall and electron inertia regimes respectively (when viewed from left to right).}
\label{fig4}
\end{figure*}

In Figs. \ref{fig1} - \ref{fig4}, the kinetic and magnetic spectra, denoted by $W_\pm$ and $M_\pm$ respectively, have been plotted as a function of $k$ (where $k := k d_i$). The `$\pm$' corresponds to the two values of $\mu$ given by (\ref{mu}). In each of the plots, we have included two vertical lines, which serve to separate the ideal ($k < 1$), Hall ($k>1$ and $k<1/d_e$) and electron inertia ($k > 1/d_e$) regions.

An inspection of (\ref{WHx}) reveals that it blows up at approximately $k=1/d_e$. This feature is not present in any of the other spectra. Hence, we observe the existence of singular behaviour in Fig. \ref{fig2} that is absent in the rest of the figures. As we have reiterated earlier, we shall consider only Figs. \ref{fig1} and \ref{fig4} to be independent and of importance, since they represent the spectra arising from the energy and generalized helicity invariants. 

In each of the figures, we have explicitly labelled certain spectral indices. The reason behind our logic is that we examine the ideal, Hall and electron inertia regimes in detail in Sec. \ref{SecESRes}, where we obtain the spectra for these limiting cases. These spectra are compared against the figures, thereby serving as a mutual consistency check. 

\section{The energy spectra of extended MHD in different regimes} \label{SecESRes}
In this Section, we shall draw upon the results of Secs. \ref{SSecLimCase} and \ref{SSecSDer}, and explicitly present the power-law scalings for the spectral energies in various regimes. 

\subsection{The ideal MHD regime} \label{SSecIdealReg}
As noted in Sec. \ref{SSecLimCase}, the ideal MHD regime is obtained in the limit $k \ll 1$ in the normalized units. In this instance, it is known that $\mu_{\pm} \rightarrow \pm 1$. Thus, we end up with the following set of relations:
\begin{equation}
\label{WE1}
W_{E_{1}}\left(k\right)\, = \,\left(2\ep_{E}\right)^{2/3} k^{-5/3}\, = \,M_{E_{1}}\left(k\right),
\end{equation}
\begin{equation}
\label{WH1}
W_{H_{1}}\left(k\right)\, = \,\left(2d^{2}_{e}\ep_{H}\right)^{2/3} k^{-1}\, = \,M_{H_{1}}\left(k\right),
\end{equation}
\begin{equation}
\label{WG1}
W_{G_{1}}\left(k\right)\, = \,\left(2\ep_{G}\right)^{2/3} k^{-1}\, = \,M_{G_{1}}\left(k\right),
\end{equation}
\begin{equation}
\label{WC1+}
W_{C_{1_{+}}}\left(k\right)\, = \,\left(2\ep_{C_{+}}\right)^{2/3} k^{-1}\, = \,M_{C_{1_{+}}}\left(k\right),
\end{equation}
\begin{equation}
\label{WC1-}
W_{C_{1_{-}}}\left(k\right)\, = \,\left(2d_e^4\,\ep_{C_{-}}\right)^{2/3} k^{-1}\, = \,M_{C_{1_{-}}}\left(k\right).
\end{equation}
The magnetic energy spectral are exactly equal to the kinetic energy spectral in each instance, since $\mu^2 = 1$. Note that in each of the above expressions, the label `$1$' denotes the ideal MHD (Alfv\'enic) limit. 

\subsection{The Hall regime} \label{SSecHallReg}
The regime where Hall effects are important (and dominant) is given by $k>1$ and $d^{2}_{e} k^{2}\ll1$. In the Hall regime, there are two values for $\mu_\pm$ that are very different, and thereby necessitate a different treatment. 

By the subscript `$2$' we shall refer to the case with the above limits and where $\mu_{+} \rightarrow k^{-1}$. In this instance, we find that 
\begin{eqnarray}
\label{WE2}
W_{E_{2}}\left(k\right)&=& \left(2\ep_{E}\right)^{2/3} k^{-5/3},\nonumber\\
M_{E_{2}}\left(k\right)&=& \left(2\ep_{E}\right)^{2/3} k^{-11/3},
\end{eqnarray}
\begin{eqnarray}
\label{WH2}
W_{H_{2}}\left(k\right)&=&\left(2d^{2}_{e}\ep_{H}\right)^{2/3} k^{1/3},\nonumber\\
M_{H_{2}}\left(k\right)&=& \left(2d^{2}_{e}\ep_{H}\right)^{2/3} k^{-5/3},
\end{eqnarray}
\begin{eqnarray}
\label{WG2}
W_{G_{2}}\left(k\right)&=& \left(2\ep_{G}\right)^{2/3} k^{1/3},\nonumber\\
M_{G_{2}}\left(k\right)&=&\left(2\ep_{G}\right)^{2/3} k^{-5/3},
\end{eqnarray}
\begin{eqnarray}
\label{WC2+}
W_{C_{2_{+}}}\left(k\right)&=& \left(2\ep_{C_{+}}\right)^{2/3} k^{-7/3},\nonumber\\
M_{C_{2_{+}}}\left(k\right)&=&\left(2\ep_{C_{+}}\right)^{2/3} k^{-13/3},
\end{eqnarray}
\begin{eqnarray}
\label{WC2-}
W_{C_{2_{-}}}\left(k\right)&=& \left(2d_e^4\,\ep_{C_{-}}\right)^{2/3} k^{1/3},\nonumber\\
M_{C_{2_{-}}}\left(k\right)&=&\left(2d_e^4\,\ep_{C_{-}}\right)^{2/3} k^{-5/3}.
\end{eqnarray}
In the other limit, we are interested in the case where $\mu_{-}\rightarrow -k$, and this case is represented by the label `$3$' henceforth. In this instance, the spectra are given by
\begin{eqnarray}
\label{WE3}
W_{E_{3}}\left(k\right)&=& \left(2\ep_{E}\right)^{2/3} k^{-3},\nonumber\\M_{E_{3}}\left(k\right)&=& \left(2\ep_{E}\right)^{2/3} k^{-1},
\end{eqnarray}
\begin{eqnarray}
\label{WH3}
W_{H_{3}}\left(k\right)&=&\left(2d^{2}_{e}\ep_{H}\right)^{2/3} k^{-7/3},\nonumber\\M_{H_{3}}\left(k\right)&=& \left(2d^{2}_{e}\ep_{H}\right)^{2/3} k^{-1/3},
\end{eqnarray}
\begin{eqnarray}
\label{WG3}
W_{G_{3}}\left(k\right)&=&\left(2\ep_{G}\right)^{2/3} k^{-7/3},\nonumber\\M_{G_{3}}\left(k\right)&=& \left(2\ep_{G}\right)^{2/3} k^{-1/3},
\end{eqnarray}
\begin{eqnarray}
\label{WC3+}
W_{C_{3_{+}}}\left(k\right)&=& 2^{-4/3} \left(2\ep_{C_{+}}\right)^{2/3} k^{-7/3},\nonumber\\
M_{C_{3_{+}}}\left(k\right)&=&2^{-4/3} \left(2\ep_{C_{+}}\right)^{2/3} k^{-1/3},
\end{eqnarray}
\begin{eqnarray}
\label{WC3-}
W_{C_{3_{-}}}\left(k\right)&=& \left(2d_e^4\,\ep_{C_{-}}\right)^{2/3} k^{-7/3},\nonumber\\
M_{C_{3_{-}}}\left(k\right)&=&\left(2d_e^4\,\ep_{C_{-}}\right)^{2/3} k^{-1/3}.
\end{eqnarray}

\subsection{The electron inertia regime} \label{SSecEIReg}
When the electron inertia effects become important, and dominate the landscape, the conditions $k\gg1$ and $d^{2}_{e} k^{2}\gg1$ must be met. In this regime, the two values of $\mu_\pm$ give rise to different spectra, as in Sec. \ref{SSecHallReg}. 

In the first case, $\mu_{+}\rightarrow 1/k$, and this is denoted by the subscript `$4$'. The various spectra exhibit the following scalings:
\begin{eqnarray}
\label{WE4}
W_{E_{4}}\left(k\right)&=& \left(2\ep_{E}\right)^{2/3} k^{-5/3},\nonumber\\M_{E_{4}}\left(k\right)&=& \left(2\ep_{E}\right)^{2/3} k^{-11/3},
\end{eqnarray}
\begin{eqnarray}
\label{WH4}
W_{H_{4}}\left(k\right)&=&\left(2d_e^{-2}\ep_{H}\right)^{2/3} k^{-7/3},\nonumber\\
M_{H_{4}}\left(k\right)&=& \left(2d_e^{-2}\ep_{H}\right)^{2/3} k^{-13/3},
\end{eqnarray}
\begin{eqnarray}
\label{WG4}
W_{G_{4}}\left(k\right)&=&\left(2d_e^{-2}\ep_{G}\right)^{2/3} k^{-7/3},\nonumber\\
M_{G_{4}}\left(k\right)&=& \left(2d_e^{-2}\ep_{G}\right)^{2/3} k^{-13/3},
\end{eqnarray}
\begin{eqnarray}
\label{WC4+}
W_{C_{4_{+}}}\left(k\right)&=& \left(2\ep_{C_{+}}\right)^{2/3} k^{-7/3},\nonumber\\
M_{C_{4_{+}}}\left(k\right)&=& \left(2\ep_{C_{+}}\right)^{2/3} k^{-13/3},
\end{eqnarray}
\begin{eqnarray}
\label{WC4-}
W_{C_{4_{-}}}\left(k\right)&=& 2^{-4/3} \left(2\ep_{C_{-}}\right)^{2/3} k^{-7/3},\nonumber\\
M_{C_{4_{-}}}\left(k\right)&=&2^{-4/3} \left(2\ep_{C_{-}}\right)^{2/3} k^{-13/3}.
\end{eqnarray}

When we consider the other limit, it corresponds to $\mu_{-}\rightarrow 1/\left(d^{2}_{e}k\right)$ and we use the label `$5$' to identify this case. The spectral distributions for the magnetic and kinetic energies are
\begin{eqnarray}
\label{WE5}
W_{E_{5}}\left(k\right)&=& \left(2d_e^2\,\ep_{E}\right)^{2/3} k^{-5/3},\nonumber\\
M_{E_{5}}\left(k\right)&=& \left(2d_e^{-4}\,\ep_{E}\right)^{2/3} k^{-11/3},
\end{eqnarray}
\begin{eqnarray}
\label{WH5}
W_{H_{5}}\left(k\right)&=& \left(2d_e^2\,\ep_{H}\right)^{2/3} k^{-7/3},\nonumber\\
M_{H_{5}}\left(k\right)&=& \left(2d_e^{-4}\,\ep_{H}\right)^{2/3} k^{-13/3},
\end{eqnarray}
\begin{eqnarray}
\label{WG5}
W_{G_{5}}\left(k\right)&=& \left(2\ep_{G}\right)^{2/3} k^{-7/3},\nonumber\\M_{G_{5}}\left(k\right)&=& \left(2\ep_{G}\right)^{2/3} d_e^{-4} k^{-13/3},
\end{eqnarray}
\begin{eqnarray}
\label{WC5+}
W_{C_{5_{+}}}\left(k\right)&=& 2^{-4/3} \left(2\ep_{C_{+}}\right)^{2/3} k^{-7/3},\nonumber\\
M_{C_{5_{+}}}\left(k\right)&=& 2^{-4/3} \left(2\ep_{C_{+}}\right)^{2/3} d_e^{-4} k^{-13/3},
\end{eqnarray}
\begin{eqnarray}
\label{WC5-}
W_{C_{5_{-}}}\left(k\right)&=& \left(2d_e^4\,\ep_{C_{-}}\right)^{2/3} k^{-7/3},\nonumber\\
M_{C_{5_{-}}}\left(k\right)&=& \left(2d_e^{-2}\,\ep_{C_{-}}\right)^{2/3} k^{-13/3}.
\end{eqnarray}
This completes our analysis of the spectra in the different regimes, and for the various choices of the parameters. Our scalings are verified to be entirely consistent with the plots presented in Sec. \ref{SSecSpecPlot}.

We wish to observe that the primary difference between our model, and the results obtained in \citet{KM04} is that the latter lacks electron inertia effects. Hence, the results of Secs. \ref{SSecIdealReg} and \ref{SSecHallReg} are identical to that of \citet{KM04}, but our results in the regime where electron inertia effects are dominant, viz. the findings of Sec. \ref{SSecEIReg}, are altogether new.

\section{Discussion and analysis} \label{SecDiscAn}
As the ideal MHD regime has been studied by many authors (see for e.g. the text by \citealt{Bisk03}), we shall not focus on it in detail. Instead, we focus primarily on the Hall and electron inertia regimes in our analysis. 

Let us commence our comparison by first studying the Hall regime, and comparing our results with the detailed analytical and numerical results of \citet{GB07}; some of the chief conclusions obtained therein were also corroborated by \citet{MG12}. The simulations undertaken by \citet{GB07} demonstrated that the magnetic fluctuations can exhibit a wide range of power-law behaviour. This conclusion matches our results in Sec. \ref{SSecHallReg}. Moreover, a careful inspection of Sec. 3.2 of \citet{GB07} confirms that their findings are in exact agreement with our model:
\begin{enumerate}
\item As per \citet{GB07}, the kinetic energy exhibits a $-5/3$ slope, whilst the magnetic energy is characterized by a $-11/3$ spectrum in the Hall regime. This is precisely the scaling obtained in (\ref{WE2}). 
\item It was found in \citet{GB07} that the magnetic energy displays a $-5/3$ scaling at large scales, and the $-11/3$ scaling at small scales. This is in contrast to the kinetic energy which displays the $-5/3$ behaviour at all scales. A careful inspection of (\ref{WE1}) and (\ref{WE2}) confirms that this is indeed the case. 
\item The fact that the magnetic energy is slightly greater than the kinetic energy can be explained naturally via Hall MHD \citep{KM04,GB07,SMC08,SP15}, and is also consistent with observations \citep{GVM91,Mar06,BC13}. The $\mu_{-}$ case in the Hall regime, that was studied in Secs. \ref{SSecLimCase} and \ref{SSecHallReg}, is consistent with these results.
\end{enumerate}
One minor difference between our results and that of \citet{GB07,MG12} that the upper bound on the magnetic energy spectral index is $-11/3$ in the latter case, whereas our model suggests that $-13/3$ can be achieved, as seen from (\ref{WC2+}). The scaling of $-13/3$ is also supported by the previous Kolmogorov-like analysis of Hall MHD by \citet{KM04} who also emphasized the important point that, in their model, the steepened spectra were very much a part of the inertial range, and were distinct from the dissipation range invoked in earlier studies.

In general, the fact that the Hall and electron inertia regimes predict slopes steeper than $-5/3$ is not a surprising fact, as this prediction has plenty of observational evidence in its favour \citep{SGL01,SGRK09,SGBCR10,BC13}. One of the remarkable features of the solar wind turbulence spectrum is the potential existence of three different magnetic spectra, with spectral `breaks' separating them \citep{BCXZ15}, of which two are well-documented: the Kolmogorov $-5/3$ spectrum at large scales, and an extended inertial range between the ion and electron gyroscales with an index of approximately $-2.5$ to $-3$ \citep{SHVL06,ACVS07,ACVS08,SGRK09,Alex09,BAMM12,FLMVH15}. The last, on the other hand, is quite contested since it has been modelled as a power-law with an index of possibly around $-4$ by \citet{SGRK09} (see also \citet{SHVL06}), and as an exponential by \citet{Alex09,ALM12}. We will return to this aspect later and examine the reasons behind this ambiguity in greater detail. 3D anisotropic spectra have also suggested that such steep power laws do exist at sufficiently small scales \citep{SGBCR10}. This range is sometimes referred to as the dissipation range, and merits an extended discussion of its features below.

If we suppose that such (steep) power laws do exist, one must search for potential candidates to explain this behaviour. At such scales, kinetic effects are likely to play an important role. For instance, it is expected that Landau damping plays a major role, in conjunction with the Kinetic Alfv\'en Wave (KAW) and (passive) ion entropy cascades, by transferring the energy to collisional scales and leading to ion and electron heating \citep{Sch09}. Collisionless damping also plays an important role in regulating the spectra in the dissipation range. For instance, it was shown in \citet{Het08} by means of a local cascade model (with critical balance) that the spectrum could exhibit an exponential fall-off, quite similar to the results obtained in \citet{Alex09,ALM12}. On the other hand, when the critical balance conjecture was abandoned, it was shown in \citet{HTD11} that steep spectra that were nearly power-law in nature could be obtained, thus analogous to the analysis of solar wind observations undertaken by \citet{SGRK09}. 

In addition, there are many other effects associated with Landau damping, as a result of which it has been identified as a major player in explaining the non-universal power law spectra of the solar wind \citep{PS15}. In addition to Landau damping, we also wish to point out the major role played by other kinetic effects such as pressure anisotropy and its accompanying kinetic instabilities \citep{Kunzet15}, phase mixing \citep{Schet16}, intermittency and coherent structures \citep{Servi15}; a summary of some of these aspects can be found in \citet{Sch09}.

Hence, there has been a great deal of work centred around (gyro)kinetic simulations of the solar wind \citep{Het08,CB11,THD13,Tet15} and hybrid fluid-kinetic models \citep{Vet07,CDQB11,Tol16}. Regardless of the physical model used, either kinetic Alfv\'enic waves or whistler waves are the primary candidates responsible for this turbulence \citep{CBet13,BHXP13}. The analysis by \citet{PBG10} suggests that the former cannot serve as a viable candidate, as the whistlers subject to collisionless damping and do not reach the electron gyroscale (see also \citealt{GS09}), but this issue cannot be said to have been conclusively settled. This opens up the possibility of using Hall MHD, which serves as a natural model for whistler waves \citep{MK05,Gal06}. Typically, Hall MHD and/or whistler turbulence yield spectra with the slope of $-7/3$ \citep{SZ05,ACVS07,GB07,ACVS08,SS09,MG12,SP15}, which falls within the second, and \emph{not} the third, range as per the observational evidence.

Furthermore, there are some inherent limitations to using Hall MHD as a universal model for solar wind turbulence, as discussed in \citet{How09}. From the perspective of two-fluid theory, the effects of electron inertia cannot be ignored at scales comparable to the electron skin depth \citep{KT73}, implying that Hall MHD cannot serve as our physical model.\footnote{It is known that $\beta \approx 1-2$ for the solar wind \citep{MS06,SGRK09}, implying that the electron gyroradius and skin depth are approximately equal to each other.} It is at this juncture that we invoke the results from Sec. \ref{SSecEIReg} that accurately capture the effects of electron inertia (as extended MHD was used in this work).

A careful scrutiny of Sec. \ref{SSecEIReg} reveals that \emph{all} of the magnetic energy spectral indices are either $-11/3$ or $-13/3$. We particularly emphasize the $-13/3$ slope as this does not appear to have been predicted before by any of the existent fluid models in the electron inertia regime, although \citet{KM04} had discussed this scaling in the context of Hall MHD earlier. It is also very intriguing to note that the theoretically predicted slope of $-13/3$ is quite close to the value of $-4.16$ that was obtained from the solar wind observations at the smallest scales \citep{SGRK09}.

A cautionary statement is necessary: although the predictions of our model are quite similar to the solar wind data, the latter cannot be viewed as exact in this regime on account of instrumental inaccuracies \citep{Set13,ACSHB13}. Instead, it has been shown that a spectrum of slopes peaked around approximately $-4$ is manifested \citep{ALM12,Set13}. Hence, we can argue that our scalings are fairly close to the experimental evidence, as well as the 2D and 3D PIC simulation studies by \citet{CB11} and \citet{GCW12} which have reported fairly similar results. We also wish to emphasize that a steep spectra, with a power-law index of $-4.228$, has also been observed for the interplanetary magnetic field, and this fact is evident upon inspection of Fig. 1 of \citet{Lea98}. This is conventionally attributed to the `dissipation' range, but it is possible that this spectra could arise from the existence of an extended inertial range that gives rise to the aforementioned $-13/3$ spectrum in the electron inertia (and Hall) regime.

The $-11/3$ slope is interesting in its own right, as it exactly matches the results from the two-fluid simulations of \citet{AGMDG14}. The $-11/3$ spectrum also arises when electron MHD is used as the basic physical model \citep{MG10}. As pointed out in Sec. \ref{SecIntro}, electron MHD is a limiting case of extended MHD, and it is founded on the narrow assumption that the ions are stationary. Thus, it is the complexity and broad scope of our model that is primarily reasonsible for recovering a diverse spectrum of results in Sec. \ref{SecESRes}. 

One of the other features that emerges from the wide-ranging nature of our model is that, in Sec. \ref{SSecEIReg} and the first half of Sec. \ref{SSecHallReg}, we find that the magnetic energy spectra differs from the kinetic energy spectra by a factor that is proportional to $1/k^2$. This arises on account of the fact that there is a factor of $\mu^2$ involved, and $\mu \propto k^{-1}$ in these instances. We observe that a somewhat similar result has been presented in \citet{BPBP11}, whose detailed analysis of MHD simulations and observations revealed that $W-M \propto k_\perp^{-2}$. 

We end our analysis on a cautionary note, by summarizing some of the limitations of our treatment. Whilst it is true that extended MHD is a much better model than ideal (or Hall) MHD, it does not capture kinetic or dissipative effects. Moreover, we have not addressed the issue of parallel vs perpendicular (with respect to the mean magnetic field) magnetic fluctuations in our analysis, and this anisotropy is known to be an important feature of the solar wind \citep{BC13} and other astrophysical plasmas \citep{Bisk03}. However, we believe that the consistency of our results with all of the previous studies described above augurs well for this model. We also find that the spectra are not truly power laws (in the universal sense), as seen from Sec. \ref{SSecSpecPlot}. This agrees with the recent overview presented in \citet{BC13}; see also the arguments put forth in \citet{Pod16}.

\section{Conclusion} \label{SecConc}
It is widely known that ideal MHD is not an appropriate theory in many space and astrophysical plasmas. Many of the limitations inherent to MHD can be bypassed by adopting extended MHD as the base model, since it encompasses two-fluid effects such as the Hall drift and electron inertia. In this paper, we utilized extended MHD to study the resultant magnetic and energy spectra by resorting to a line of reasoning akin to the one adopted in \citet{Kol41}.

We studied the properties of nonlinear Alfv\'en waves in extended MHD, which built upon the earlier works by \citet{KM04,MK05,MM09}; an alternative derivation of the same results, from a more mathematical perspective, can be found in \citet{AY16}. The primary results of this analysis can be summarized by (\ref{bvRel}) and (\ref{mu}), which yielded exact relations between the magnetic and kinetic fluctuations in Fourier space. We also presented the invariants of extended MHD, which comprised of the energy and two generalized helicities that possessed the same mathematical structure as the magnetic or fluid helicity.

These results were employed in conjunction with a Kolmogorov-based argument, namely that the cascade rates of the energy and generalized helicities (taken to be constants) were a product of the eddy turnover time and the corresponding energy and generalized helicities. We demonstrated that this procedure led us to the kinetic and magnetic energy spectra, which were presented and plotted in Sec. \ref{SecSpecDis}. This was followed by a detailed analysis of the spectra in the ideal $\left(k < 1/\lambda_i\right)$, Hall $\left(1/\lambda_i < k < 1/\lambda_e \right)$, and electron inertia $\left(k > 1/\lambda_e\right)$ regimes. 

The chief differences from previous works stem from our consideration of the electron inertia regime, which was missing in previous Hall MHD based studies, such as \citet{KM04,Gal06,GB07}. By and large, we find that most of the spectra exhibit steepening in the electron inertia regime. In the Hall regime, the spectral index of $-11/3$ and $-13/3$ occur only once for the magnetic spectra, as seen from Sec. \ref{SSecHallReg}. On the other hand, Sec. \ref{SSecEIReg} reveals that the electron inertia regime is characterized \emph{only} by these two slopes for the magnetic spectra. More precisely, a comparison of the expressions for $W_{C_{-}}$ and $M_{C_{-}}$ in the Hall and electron inertia regimes reveals that the spectral indices are $+1/3$ and $-5/3$ (for $\mu_+$) in the former, and $-7/3$ and $-13/3$ in the latter. This is a clear manifestation of the steepening that occurs in the electron inertia regime, which cannot be captured by using Hall MHD, or even electron MHD \citep{MG10}, as the physical model.

We undertook a detailed comparison of our results in Sec. \ref{SSecHallReg} with other theoretical and numerical studies involving Hall MHD, and established the veracity of our results in Sec. \ref{SecDiscAn}. We followed this up with a general comparison against observational and numerical studies of the solar wind, which has proven to be an excellent means of testing different fluid and kinetic theories of turbulence. One of the chief (if somewhat contested) observational findings from the solar wind is the existence of a cascade at length scales smaller than the electron skin depth \citep{SGRK09}. The magnetic spectrum has been argued to exhibit a power law with an index of approximately $-4$, which is very close to our theoretical predictions of either a $-11/3$ or $-13/3$ slope in the electron inertia regime; the derivation of these scalings can be found in Sec. \ref{SSecEIReg}. Moreover, we have also pointed out the closeness of the $-13/3$ spectrum to measurements of the interplanetary magnetic field \citep{Lea98}, where a slope of $-4.228$ was obtained. A crucial distinction between our work and other theoretical methodologies in the literature is that the results in this paper rely upon the assumption of an (extended) inertial range, as opposed to the usual concept of the `dissipation' range at these small scales.

Hence, the tentative agreement with the observational results lends some credence to the fact that the (theoretically predicted) scalings may emerge from the presence of an extended inertial range, as opposed to the dissipation range - a hypothesis that was proposed in the Hall MHD based turbulent model of \citet{KM04}. We also compared our work with numerical studies \citep{AGMDG14} involving electron inertia effects, and showed that the $-11/3$ spectrum thus obtained is consistent with our findings. Moreover, as electron MHD is a subset of extended MHD \citep{KLMWW14}, we have verified that previous results derived from the former \citep{MG10} are duly recovered by using the latter model. 

As the observational and numerical scalings determined in the electron inertia regime are in good agreement with our theoretical findings, we suggest that extended MHD constitutes a viable model for extracting the turbulent spectra across all scales, including those smaller than the electron gyroradius. We end this work on a cautionary note, by pointing out the fact that anisotropy, compressibility, dissipation, and kinetic effects (such as Landau damping) have not been incorporated into our analysis. It is our intention to make them the subject of our future investigations.

\acknowledgments
HMA would like to thank the Egyptian Ministry of Higher Education for supporting his research activities, and gratefully acknowledges the hospitality of the Princeton Plasma Physics Laboratory where this work was initiated. ML was supported by the DOE (Grant No. DE-AC02-09CH-11466) and the NSF (Grant No. AGS-1338944) whilst pursuing this work. SMM's work was supported by the DOE (Contract No.DE-FG03-96ER-54366). The authors wish to thank the referee for the helpful comments and suggestions, which helped improve the quality of the paper.

\bibliographystyle{aasjournal} 
\bibliography{XMHD}

\end{document}